\documentclass[reprint,superscriptaddress,amsmath,amssymb,aps,prd,
showkeys,showpacs,nofootinbib] {revtex4-1}
\usepackage[utf8x]{inputenc}
\usepackage{graphicx}
\usepackage{bm}
\usepackage{hyperref}
\usepackage{dcolumn}
\usepackage{slashed}

\begin{document}
\title{Renormalization of masses of sterile neutrinos in the $\nu$MSM}
\author{Ananda Roy}
\email{ananda.roy@epfl.ch}
\affiliation{Indian Institute of Technology Kanpur, Kanpur, Uttar
Pradesh 208016,India}
\affiliation{Institut de Théorie des Phénomènes Physiques, EPFL,
CH-1015 Lausanne, Switzerland}

\begin{abstract}
The quasi-degeneracy of heavier sterile neutrino masses in the
Neutrino Minimal Standard Model ($\nu$MSM) facilitates the production
of lepton asymmetry below the electroweak scale. The first
order loop corrections to this mass-difference has been computed in
this work along with a numerical estimate of the contribution. 
\end{abstract}
\keywords{$\nu$MSM, loop corrections, mass difference, neutrino
physics}
\pacs{11.10.Hi,12.10.-g,12.60.-i,13.15.+g}
\maketitle

\section{Introduction}

A renormalizable extension of the Standard Model (SM), the $\nu$MSM
extends the existing SM of particle physics by three sterile
neutrinos, which are singlets under the SM gauge group. The $\nu$MSM
attempts to address the various unresolved issues of the SM
\cite{Asaka:2005an,Asaka:2005pn,Shaposhnikov:2008pf,Bezrukov:2007ep}
(for a review see \cite{Boyarsky:2009ix}), one of them being
dark matter production.  The lepton asymmetry, produced at the same
time as the generation of the dark matter sterile neutrinos (at
temperature of $\sim 100$ MeV),  affects their spectrum and number
density (see \cite{Shi:1998km,Asaka:2006nq,Laine:2008pg}). Upon
comparison of theoretical computation of the abundance of dark matter
sterile neutrino with cosmological and astrophysical observations, the
lepton asymmetry ($\Delta L$) is required to be much larger than
the baryon asymmetry ($\Delta$B): $\frac{\Delta L}{\Delta
B}\geq3\times10^5$, where $\Delta B\sim 10^{-10}$.
\footnote{Different  possible ways to modify this requirement may be
found in \cite{Shaposhnikov:2006xi,Kusenko:2006rh,Anisimov:2008qs,
Bezrukov:2009yw,Bezrukov:2008ut}.}

 The high value of the lepton-asymmetry can be created provided the
masses of the heavier sterile neutrinos in the $\nu$MSM are almost
degenerate.
Moreover, the mass-difference that leads to the requisite value of the
lepton asymmetry must be much lower than the active neutrino
mass-difference \cite{Shaposhnikov:2008pf,ananda2}. In an
accompanying paper \cite{ananda2}, we discussed the naturalness of
the fine-tuning required for satisfying the leptogenesis and
active neutrino oscillation observations from the perspective of the
renormalization group (RG) evolution of the $\nu$MSM parameters. A 
complete analysis of the effect of mass-difference on leptogenesis,
however, requires the
computation of the physical mass-difference. This is the purpose
of the present work, where we compute the radiative corrections to
the mass-difference at the one loop level.  For a review
of the generalized
on-shell renormalization procedure for Majorana neutrino theories, see
\cite{Almasy:2009kn}.

The paper is organized as follows: in section \ref{sec2}, we review
the  Lagrangian of the $\nu$MSM, along with the definitions of the
different parameters. In section \ref{sec4}, we describe our
formalism of computation of loop corrections and compute the various
contributions. In section \ref{sec5}, we compute the mass difference
with the loop corrections taken into account, while numerical
estimates are given in \ref{sec6}. Finally,
in section \ref{sec7}, we summarize our results. 

\section{The $\nu$MSM Lagrangian and Relevant Mass-Matrix}\label{sec2}
We use the Lagrangian of the $\nu$MSM in the parametrization
\cite{Shaposhnikov:2008pf,ananda2,Shaposhnikov:2006nn}. In addition,
considering the fact that the sterile neutrino Yukawa couplings are
much smaller than the gauge couplings and neglecting the numerically
much smaller charged lepton Yukawa couplings, we choose a basis for
the leptonic doublets, in which the Lagrangian has the following
simple form: 
\begin{equation}
 {\cal L}_{\nu MSM}={\cal L}_0+\Delta{\cal L},
\end{equation}
\begin{eqnarray*}
 {\cal L}_0={\cal L}_{SM}+
 \sum_{I=2,3}\overline{N_I}i\partial_\mu\gamma^\mu N_I
 +(f_2\overline{l_2}N_2
\end{eqnarray*}
\begin{equation}\label{lag1}
 \hspace{20mm}+f_3\overline{l_3}N_3)\tilde{\phi}-M\overline{N_2^c}
N_3+h.c.,
\end{equation}
\begin{equation}\label{lag2}
 \Delta{\cal L}=f_{23}\overline{l_2}
N_3\tilde{\Phi} -\frac{\Delta
M}{2}\sum_{I=2,3}\overline{N_I^c} N_I + h.c.,
\end{equation}
where $N_I$ are the right handed singlet leptons ($I=2,3$), $\phi$ and
$l_{2,3}$ are the Higgs and the lepton doublets
respectively, $M$ is
the common mass of the two heavy neutral fermions, $\Delta M$ is the
diagonal element of the Majorana mass matrix,
$\tilde{\phi}_i=\epsilon_{ij}\phi^*_{j}$, M and $\Delta M$ are taken
to be real. The Yukawa couplings $f_2,f_3$ can be
chosen to be real by suitably defining the phases of $l_2,l_3$, while
the $f_{23}$ is complex, with a phase of $n$. The relation between
$f_2,f_3$ and $h_{\alpha I}$ introduced in
\cite{Shaposhnikov:2008pf,ananda2} can be found by comparing eqs. 
\eqref{lag1} and \eqref{lag2} with eqs. (2) and (3) of
\cite{ananda2}. 

 We have omitted the dark matter sterile neutrino $N_1$ from the
Lagrangian as its influence on the problem we are interested in is
negligibly small \cite{Shaposhnikov:2008pf}.

We will use unitary gauge ($\Phi=\frac{1}{\sqrt{2}}(0\ v+h)^T$) to
reduce the computational complexity. Here $v$ is the vacuum
expectation value of the Higgs boson and is taken to be 246 GeV. For
a list of conventions used
for the computation, see appendix \ref{notcon}.

Using the Euler-Lagrange's equation of motion, we get the Dirac
equation for the system of particles as follows:
\begin{equation}
 i\slashed{\partial}\begin{pmatrix}
	                    N_{2R} \\
			    N_{3R} \\
			    \nu_{2L}^c \\
 			    \nu_{3L}^c \\
			    N_{2R}^c \\
    			    N_{3R}^c \\
			    \nu_{2L} \\
			    \nu_{3L} 
	                   \end{pmatrix} -\begin{pmatrix}
					0	&	M_\nu\\
					M_\nu^\dagger	&	0
					\end{pmatrix}\begin{pmatrix}
	                    N_{2R} \\
			    N_{3R} \\
			    \nu_{2L}^c \\
 			    \nu_{3L}^c \\
			    N_{2R}^c \\
    			    N_{3R}^c \\
			    \nu_{2L} \\
			    \nu_{3L} 
	                   \end{pmatrix}=0,
\end{equation}
where \begin{equation}
        M_\nu=\begin{pmatrix}
               \Delta M	&	M	&	\frac{-vf_2}{\sqrt{2}}
&	0\\
		M	&	\Delta M	&
\frac{-vf_{23}^*}{\sqrt{2}}	&	\frac{-vf_3}{\sqrt{2}}\\
		\frac{-vf_2}{\sqrt{2}}	&
\frac{-vf_{23}^*}{\sqrt{2}}	&	0	&	0\\
		0	&	\frac{-vf_3}{\sqrt{2}}	&	0
&	0
              \end{pmatrix}.
       \end{equation}

 At the tree level, the eigenvectors of $M_\nu^\dagger M_\nu$ up
to first order in Yukawa couplings with eigenvalues of $M_\nu$ up to
second order in the same are obtained by perturbative computation and
are listed in the appendix \ref{mass_eigen} (see \cite{Justine}).

The mass-difference between the sterile neutrino flavors
were shown to be \cite{Shaposhnikov:2008pf}: 
\begin{equation}
 \delta m_{tree}=\frac{|m^2|}{M},
\end{equation}
where

\begin{equation}
 m^2\equiv f_2f_{23}v^2+2M\Delta M\label{eq2}. 
\end{equation}
The aim of the paper is to compute the first order loop corrections to
this mass difference.

Also, the active neutrino mass-difference ($\Delta m_\nu$) may be
solved at the tree-level to be \cite{Shaposhnikov:2008pf}: 
\begin{equation}\label{active}
 \Delta m_\nu=\frac{f_2|f_{23}|v^2}{M}.
\end{equation}

\section{Loop Corrections}\label{sec4}
\subsection{Propagator}

Consider the propagator for the system of active and sterile
neutrinos: 
\begin{equation}\label{prop}
 S_F(p)=\frac{i}{\slashed{p}-{\cal M}- \Sigma},
\end{equation}
where $\Sigma$ represents the first order loop correction to the mass
matrix
${\cal M}$. It is useful to keep in mind that ${\cal M}$ and $\Sigma$
represent the mass-matrix and the loop corrections of the complete
system of 2 active and 2 sterile neutrinos, and hence are $8\times 8$
matrices. 
As usual, mass eigenvalues are given at the one loop level by the
poles of the propagator.

Including only one particle irreducible diagrams for the computation,
it is easy to see that only the following loop correction matrix
elements are relevant: $\Sigma_{ss},\Sigma_{as},\Sigma_{sa}\
\text{and}\ \Sigma_{aa}$,where
\begin{equation}
 \Sigma_{ss}=\begin{pmatrix}
              \langle N_{2R}|\Sigma|N_{2R}\rangle&\langle
N_{2R}|\Sigma|N_{3R}\rangle\\\langle
N_{3R}|\Sigma|N_{2R}\rangle&\langle N_{3R}|\Sigma|N_{3R}\rangle
             \end{pmatrix}
\end{equation}
and so on. Here the subscripts $s,a$ represent sterile and active
flavors
respectively.

\subsection{Loop correction to active neutrino propagators:\\
Computation of $\Sigma_{aa}$}\label{Sigma_aa}
Considering diagrams that give up to quadratic contribution in Yukawa
coupling, we find that the only possible contribution comes from the
internal W and Z boson loops. The following figure (Fig.
\ref{active}) represents the loop-correction to the active neutrino
propagator.

\begin{figure}[htp]
 \centering
\includegraphics[height=20mm,width=40mm]{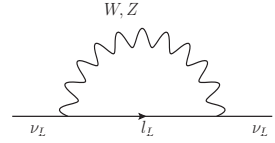}
\caption{Contribution to active neutrino propagator from W,Z boson
loop.}
\label{active}
\end{figure}
Performing the one-loop computation using dimensional
regularization, we find that

\begin{eqnarray*}
 \langle\nu_{2L}|\Sigma|\nu_{2L}\rangle=\frac{\slashed{p}}{(4\pi)^2}
\Big[-\frac{p^2}{v^2}(1+\ln\frac{\mu^2}{M_W^2})-\frac{p^2}{2v^2}
(1+\ln\frac{\mu^2}{M_Z^2})
\end{eqnarray*}
\begin{equation}
 \hspace*{20mm}+\frac{3g^2+g'^2}{24}\Big].
\end{equation}
Define: 
\begin{equation}
 C_1=\frac{1}{(4\pi)^2}\Big[-\frac{p^2}{v^2}(1+\ln\frac{\mu^2}{M_W^2}
)-\frac{p^2}{2v^2}(1+\ln\frac{\mu^2}{M_Z^2})+\frac{3g^2+g'^2}{24}\Big]
.
\end{equation}
Since the active neutrino masses are much smaller than the Higgs
vacuum expectation value, we can assume that $ p^2\ll v^2$, implying
\begin{equation}\label{eqc1}
 C_1=\frac{1}{(4\pi)^2}\frac{3g^2+g'^2}{24}.
\end{equation}

Thus,
\begin{equation}
 \langle\nu_{2L}|\Sigma|\nu_{2L}\rangle=\slashed{p}C_1.
\end{equation}
Analogous computation shows:
\begin{equation}
 \langle\nu_{3L}|\Sigma|\nu_{3L}\rangle=\slashed{p}C_1.
\end{equation}
In absence of terms contributing to the mixing between
$\nu_2$ and $\nu_3$, the lowest order contribution to the
loop correction matrix may be written as follows:  
\begin{equation}
\Sigma_{aa}=\slashed{p}C_1\begin{pmatrix}
                           1&0\\0&1
                          \end{pmatrix}.
\end{equation}
\subsection{Loop Correction to the Sterile Neutrino propagators:\\
Computation of $\Sigma_{ss}$}
The sterile neutrino propagator receives contributions from the
internal Higgs loop in addition to W and Z boson
loops. From here on, we use the intuitive notation that
$\Sigma_{w,z,h,t}$ respectively represent the contribution to the
loop-correction $\Sigma$ from W, Z, Higgs-boson loop and tadpole
graphs. 
\subsubsection{Contribution from the W boson loop}
\begin{figure}[htp]
 \centering
\includegraphics[height=20mm,width=50mm]{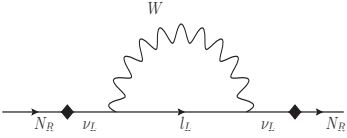}
\caption{Contribution to sterile neutrino propagator from W boson
loop; N, $\nu$ represent the different possible flavors.}
\label{sterile_w}
\end{figure}
The contribution coming from the W boson loop is represented in Fig.
\ref{sterile_w} and is given by
\begin{equation}
 \langle
N_{2R}|\Sigma_w|N_{2R}\rangle=\frac{f_2^2}{(4\pi)^2}\slashed{p}\Big[
-\frac { 1 }{2}(\ln\frac{\mu^2}{M_W^2}+1)+\frac{g^2v^2}{24p^2}\Big].
\end{equation}

\subsubsection{Contribution from Z boson loop}
\begin{figure}[htp]
 \centering
\includegraphics[height=20mm,width=50mm]{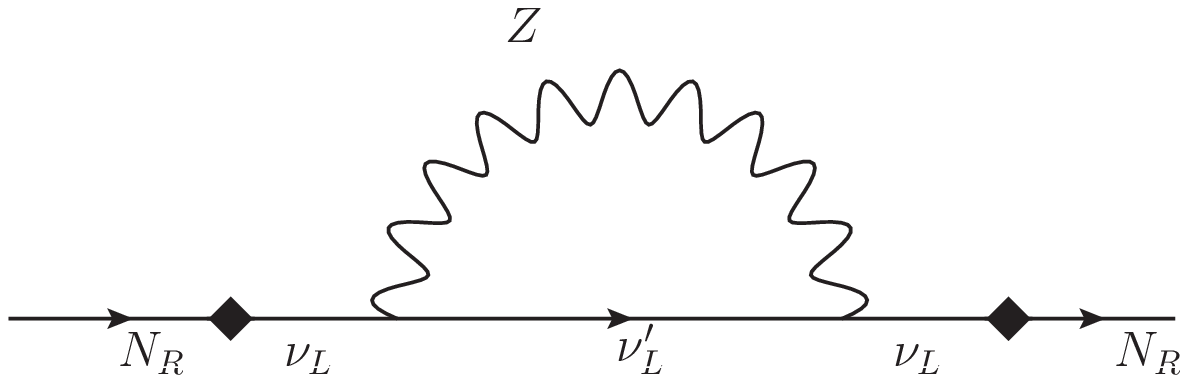}
\caption{Contribution to sterile neutrino propagator from Z boson
loop.}
\label{sterile_z}
\end{figure}
Fig. \ref{sterile_z} represents the contribution from the Z-boson
loop. Performing the computation, we get
\begin{equation}
\langle
N_{2R}|\Sigma_z|N_{2R}\rangle=\frac{f_2^2}{(4\pi)^2}\slashed{p}\Big[
-\frac { 1
}{4}(\ln\frac{\mu^2}{M_Z^2}+1)+\frac{(g^2+g'^2)v^2}{48p^2}\Big].
\end{equation}
\newpage
\subsubsection{Contribution from the Higgs boson loop}
\begin{figure}[htp]
 \centering
\includegraphics[height=20mm,width=50mm]{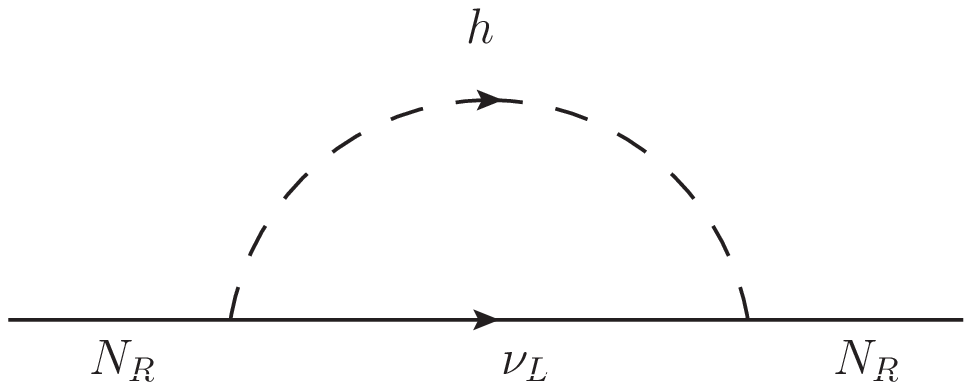}
\caption{Contribution to sterile neutrino propagator from Higgs boson
loop.}
\label{sterile_h}
\end{figure}
The above figure (Fig. \ref{sterile_h}) shows the contribution from
the Higgs loop, which leads to
\begin{equation}
 \langle
N_{2R}|\Sigma_h|N_{2R}\rangle=\frac{-f^2\slashed{p}}{2(4\pi)^2}\Big[
\frac { 1 } {2}\ln\frac{\mu^2}{m_H^2}+\frac{1}{4}\Big].
\end{equation}

\subsubsection{Matrix Element representing the loop contribution}
The total loop-correction will be the sum of the aforementioned
contributions. Thus,
\begin{equation}
 \Sigma_{ss}=\slashed{p}(\tilde{F_1}+\tilde{F_2})\begin{pmatrix}
                                 
f^2&f_2f_{23}^*\\f_2f_{23}&f_3^2+|f_{23}|^2
                                 \end{pmatrix},
\end{equation}

where 
\begin{eqnarray}\label{eqf1}
 \tilde{F_1}=\frac{-1}{2(4\pi)^2}\Big[\frac{1}{2}\ln\frac{\mu^2}{m_H^2
} +\frac {
1}{4}\Big],
\end{eqnarray}
\begin{equation*}
 \tilde{F_2}=\frac{1}{(4\pi)^2}\Big[-\frac{1}{4}(\ln\frac{\mu^2}{M_Z^2
}
+1)+\frac{(g^2+g'^2)v^2}{48p^2}\Big]
\end{equation*}
\begin{equation}\label{eqf2}
 \hspace{20mm}+\frac{1}{(4\pi)^2}\Big[-\frac{1}{
2}(\ln\frac{\mu^2}{M_W^2}+1)+\frac{g^2v^2}{24p^2}\Big].
\end{equation}

\subsection{Loop Correction to the Active-Sterile Neutrino propagator:
Computation of $\Sigma_{as}$ and $\Sigma_{sa}$}
The active-sterile neutrino propagator receives similar contribution
from the W and Z boson loops on the external active neutrino leg and
in addition receives contributions from tadpoles. We
consider only the top quark loop among the possible fermion loops in
the tadpole graphs.
\newpage
\subsubsection{Contribution from W boson}
\begin{figure}[htp]
 \centering
\includegraphics[height=23mm,width=50mm]{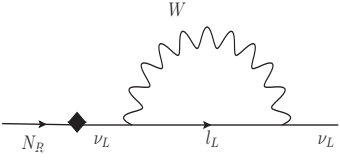}
\caption{Contribution to active-sterile neutrino propagator from W
boson loop.}
\label{as_w}
\end{figure}
The W-boson loop contribution is represented by Fig. \ref{as_w} and
it is given by
\begin{equation}
 \langle
\nu_{2L}|\Sigma_w|N_{2R}\rangle=-\frac{f_2v}{\sqrt{2}}\frac{1}{
(4\pi)^2 }
\Big[-\frac{p^2}{v^2}(1+\ln\frac{\mu^2}{M_W^2})+\frac{g^2}{12}\Big].
\end{equation}

\subsubsection{Contribution from Z boson loop}
\begin{figure}[htp]
 \centering
\includegraphics[height=23mm,width=50mm]{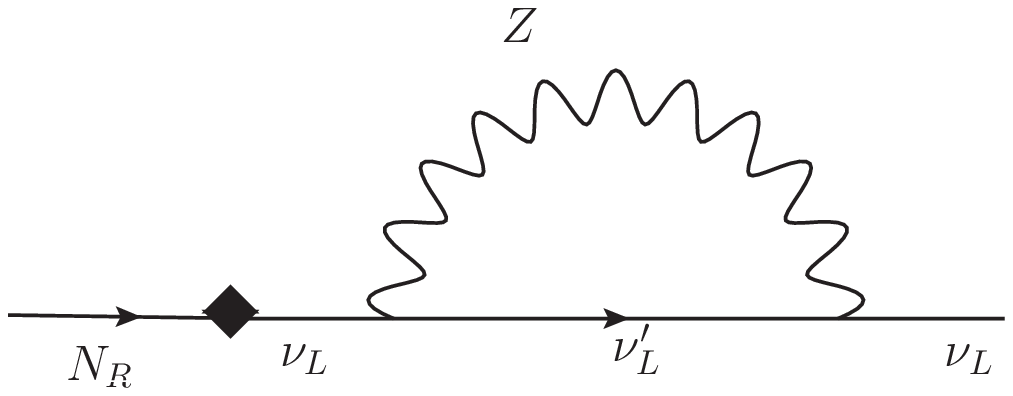}
\caption{Contribution to active-sterile neutrino propagator from Z
boson loop.}
\label{as_z}
\end{figure}
Fig. \ref{as_z} represents the contribution from the Z boson loop,
which leads to
\begin{equation}
 \langle\nu_{2L}|\Sigma_z|N_{2R}\rangle=-\frac{f_2v}{\sqrt{2}}\frac{1}
{ (4\pi)^2}\Big[-\frac{p^2}{2v^2}(1+\ln\frac{\mu^2}{M_Z^2})+\frac{
g^2+g'^2}{24}\Big].
\end{equation}
\subsubsection{Contribution from tadpole graphs}
\begin{figure}[htp]
 \centering
\includegraphics[height=28mm,width=35mm]{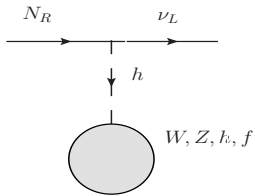}
\caption{Contribution to active-sterile neutrino propagator from
tadpole graphs.}
\label{as_t}
\end{figure}
The tadpole contribution to the active-sterile neutrino propagator is
represented by Fig. \ref{as_t}, which leads to
\begin{equation*}
\langle\nu_{2L}|\Sigma_t|N_{2R}\rangle=\frac{f_2}{(4\pi)^2}\Big[-\frac
{
3m_H^2}{2\sqrt{2}v}\ln\frac{\mu^2}{m_H^2}-\frac{6M_W^4}{\sqrt{2}vm_H^2
}\ln\frac{\mu^2}{M_W^2}
\end{equation*}
\begin{equation}
 \hspace{30mm}-\frac{3M_Z^4}{\sqrt{2}vm_H^2}\ln\frac{\mu^2}{
M_Z^2}+\frac{12m_f^4}{\sqrt{2}m_H^2v}\ln\frac{\mu^2
} { m_f^2}\Big].
\end{equation}

\subsubsection{Matrix Element representing the loop contribution}
Thus, 
\begin{equation}
 \Sigma_{as}=(K_1+K_2)\begin{pmatrix}
                       f_2&f_{23}^*\\0&f_3
                      \end{pmatrix}=(K_1+K_2)C^\dagger,
\end{equation}
\begin{equation}
 \Sigma_{sa}=(K_1+K_2)C.
\end{equation}
where \begin{equation}
       C=\begin{pmatrix}
          f_2&0\\f_{23}&f_3\\
         \end{pmatrix},
      \end{equation}
\begin{eqnarray*}
 K_1=\frac{1}{(4\pi)^2}\Big[-\frac{3m_H^2}{2\sqrt{2}v}\ln\frac{\mu^2}{
m_H^2 }-\frac{6M_W^4}{\sqrt{2}vm_H^2}\ln\frac{\mu^2}{M_W^2
}
\end{eqnarray*}
\begin{equation}\label{eqk1}
 \hspace*{20mm}-\frac{3M_Z^4}{\sqrt{2}vm_H^2}\ln\frac{\mu^2}{M_Z^2
}+\frac{12m_f^4}{\sqrt{2}m_H^2v}\ln\frac{\mu^2}{m_f^2}\Big]
\end{equation}
and 
\begin{equation*}
 K_2=-\frac{v}{\sqrt{2}}\frac{1}{(4\pi)^2}\Big[-\frac{p^2}{v^2}
(1+\ln\frac{\mu^2}{M_W^2})+\frac{g^2}{12}
\end{equation*}
\begin{equation}\label{eqk2}
 \hspace{20mm}-\frac{p^2}{2v^2}(1+\ln\frac{
\mu^2}{M_Z^2})+\frac{g^2+g'^2}{24}\Big].
\end{equation}

Reasoning as in \ref{Sigma_aa}, we arrive at $p^2\ll
v^2$. This implies
\begin{equation}
 K_2= -\frac{v}{\sqrt{2}}\frac{1}{(4\pi)^2}\frac{3g^2+g'^2}{24}.
\end{equation}

\section{Computation of the mass-difference}\label{sec5}
Having obtained the one loop corrections to the propagator of the
system of active and sterile neutrinos, we proceed to compute the mass
difference between the two heavy sterile neutrinos. 

The loop correction can be written in the matrix form as follows: 
\begin{widetext}
\begin{equation}
 \Sigma=\begin{pmatrix}
         (\tilde{F_1}+\tilde{F_2})A\slashed{p}&0&0&(K_1+K_2)C\\
	 0&\slashed{p}C_1&(K_1+K_2)C^T&0\\
	 0&(K_1+K_2)C^*&\slashed{p}(\tilde{F_1}+\tilde{F_2})A^*&0\\
	 (K_1+K_2)C^\dagger&0&0&\slashed{p}C_1
        \end{pmatrix}.
\end{equation}
\end{widetext}
In order to obtain the mass-eigenvalues, we find the poles of the
propagator given in \eqref{prop} i.e. the roots of 
\begin{equation}\label{prop1}
 \slashed{p}-{\cal M}-\Sigma=0,
\end{equation}
where
\begin{equation}
 {\cal M}=\begin{pmatrix}
           0&0&M_R&-\frac{v}{\sqrt{2}}C^*\\
	   0&0&-\frac{v}{\sqrt{2}}C^\dagger&0\\
	   M_R&-\frac{v}{\sqrt{2}}C&0&0\\
	   -\frac{v}{\sqrt{2}}C^T&0&0&0\\
          \end{pmatrix}.
\end{equation}
Considering only terms up to second order in gauge and Yukawa
couplings, we arrive at equation of the form: 
\begin{equation}
 \slashed{p}\begin{pmatrix}
             1_F&0\\0&1_F
            \end{pmatrix}-{\cal M}-\begin{pmatrix}
				    0&Q_1\\Q_2&0\end{pmatrix}=0,
\end{equation}
where $\begin{pmatrix}
         1_F&0\\0&1_F
        \end{pmatrix}$ is the unit flavor matrix and $Q_1,Q_2$ are
obtained by multiplying equation \eqref{prop1} with appropriate
matrices. 

We know that if $|e\rangle$ is an eigenvector of $M_\nu^\dagger
M_\nu$, then $\frac{1}{\sqrt{2}}\begin{pmatrix}                  
   |e^*\rangle\\|e\rangle                       
        \end{pmatrix}$ is eigenvector for the matrix $\begin{pmatrix}
0&M_\nu\\M_\nu^\dagger&0
\end{pmatrix}$.
Let $\lambda_i,i=1\cdots4$ be the eigenvalues for $\slashed{p}$ at the
tree level. Then, applying first order perturbation theory, we
obtain
\begin{equation}
 \delta\lambda_i=\frac{1}{2}\begin{pmatrix}
                  \langle e_i^*|&\langle e_i|
                 \end{pmatrix}\begin{pmatrix}
			      0&Q_1\\Q_2&0\end{pmatrix}\begin{pmatrix}
						   
|e_i^*\rangle\\e_i\rangle\end{pmatrix}.
\end{equation}
Thus,\begin{equation}
      \slashed{p}_i=\lambda_i+\delta\lambda_i,
     \end{equation}
where, $\delta\lambda_i$ represents the change in $\lambda_i$ due to
the perturbation and is a function of $p^2$. For the on-shell
computation of the mass-difference, we can safely replace $p^2$ by
$M^2$ up to second order in Yukawa couplings.  

Denoting the physical mass-difference as $\delta m_{phys}$, we arrive
at a relation: 
\begin{equation}
 \delta m_{phys}=\delta m_{tree}+\delta m_{loop},
\end{equation}

\begin{eqnarray}
 \Rightarrow{\delta m}_{phy}&=&\Big|2\Delta
M+\frac{f_2f_{23}v^2}{M}\Big|+{\delta m}_{loop}.
\end{eqnarray}
where neglecting terms of order $C_1^2$,
\begin{widetext}
\begin{eqnarray*}
 {\delta m}_{loop}=\Big[\Delta M \cos
a(f_2^2+f_3^2+|f_{23}|^2)+2f_2f_{23}M\cos(a+n)\Big](\tilde{F_1}
+\tilde{F_2})
\end{eqnarray*}
\begin{equation}\label{deltamloop}
 \hspace*{15mm}+\frac{f_2f_{23}v^2}{M}\cos(a-n)
C_1-\frac{2\sqrt{2}f_2f_{23}v}{M}\cos n\cos a\ (K_1+K_2).
\end{equation}
\end{widetext}
This is central result of this work. As expected, the physical
mass-difference $\delta m_{phys}$ is RG invariant,
which can be verified using the RG equations for the parameters
involved. The necessary RG equations are given, for example, in
\cite{ananda2}.\footnote{The conventions for the scalar quartic
coupling are different in the two papers, which must be kept in mind
while performing the verification.} 
\section{Numerical Estimates}\label{sec6}
In this section, we provide a numerical estimate of the loop
contribution to the mass-difference. Evaluating equation
\eqref{deltamloop} at $\mu\sim M_W$, we obtain that 
\begin{equation}
 \delta m_{loop}(M_W)\simeq \frac{2f_2f_{23}v^2}{(4\pi)^2M}.
\end{equation}
 Comparing with our result with \eqref{active}, we see
that
\begin{equation}
 \delta m_{loop}(M_W)\sim \Delta m_\nu,
\end{equation}
where $\Delta m_\nu$ is the active neutrino mass-difference. 
Thus, the loop correction may be absorbed into  the Higgs condensate
contribution to the tree-level mass-difference. The numerical
estimate of the loop correction is taken into consideration for the
fine-tunings made on the parameters of the $\nu$MSM to satisfy the
leptogenesis conditions, as is described in an accompanying paper (see
\cite{ananda2}). 

\section{Conclusions}\label{sec7}

The generation of low temperature lepton-asymmetry has considerable
impact on the production of dark matter sterile neutrinos.
Oscillations or decays of the singlet neutrinos in the $\nu$MSM can
give rise to the requisite lepton asymmetry provided their masses are
sufficiently degenerate. A complete analysis of the problem requires
computation of the physical mass-difference between the heavier
neutrinos. In this paper, we computed the loop corrections to this
mass-difference. On performing the computation, we see that the loop
correction is of the same order as the active neutrino
mass-difference. The loop-correction is then incorporated into the
considerations for satisfying the leptogenesis conditions, which is
done in an accompanying paper \cite{ananda2}. 
\begin{acknowledgments}

The work was facilitated by the Student Exchange program between
Ecole Polytechnique Federale de Lausanne and Indian Institute of
Technology, Kanpur. The author would like to thank Mikhail
Shaposhnikov for suggesting the problem and extremely helpful
discussions. 
\end{acknowledgments}

\appendix

\section{Notations and Conventions}\label{notcon}
\subsection{Majorana Fermions}
\begin{equation}
 \gamma^\mu=\begin{pmatrix}
             0&\sigma^\mu\\\overline{\sigma}^\mu&0
            \end{pmatrix},
\gamma_5=\begin{pmatrix}1&0\\0&-1\end{pmatrix}.
\end{equation}
\begin{equation}
 \sigma^\mu=(1,{\bm\sigma}),\overline{\sigma}^\mu=(1,-{\bm\sigma}). 
\end{equation}
where the Pauli matrices are given by the standard representation
\begin{equation}
 \sigma^1=\begin{pmatrix}
           0&1\\1&0
         
\end{pmatrix},\sigma^2=\begin{pmatrix}0&-i\\i&0\end{pmatrix},
\sigma^3=\begin{pmatrix}1&0\\0&-1\end{pmatrix}.
\end{equation}
The left and right projectors are given by:
\begin{equation}
 P_L=\frac{1}{2}(1+\gamma_5), P_R=\frac{1}{2}(1-\gamma_5). 
\end{equation}
\begin{equation}
 \psi^c=C\overline{\psi}^T, C^\dagger=C^{-1},C^T=-C,
C^\dagger\gamma_\mu C=-\gamma_\mu^T.
\end{equation}
In our convention, the sterile neutrinos are Majorana particles with
the Dirac spinor written as: 
\begin{equation}
 N=\begin{pmatrix}
  N_{R}^c\\N_R
 \end{pmatrix}.
\end{equation}
Thus, $P_L N=N_R^c$ and $P_R N=N_R$. 
For the active neutrinos, the right handed components do not enter
into the Lagrangian. 
So, we can define Dirac spinor for the active neutrino as follows: 
\begin{equation}
 \nu=\begin{pmatrix}
	\nu_L\\\nu_R
     \end{pmatrix}.
\end{equation}
Using the properties of Charge conjugation and the commutation rules
for the $\gamma_5$, we get, 
\begin{equation}
 \nu_L^c=P_R\nu^c, \overline{N_R^c}=\overline{N^c}P_R.
\end{equation}
Lastly, $N^c=N$, in accordance with the fact that N are Majorana
particles.
\subsection{Higgs sector}\label{higgs}
We use the following form of the Higgs potential: 
\begin{equation}
 V(\phi)= -\mu^2(\phi^\dagger \phi)+\lambda (\phi^\dagger\phi)^2. 
\end{equation}
The vacuum expectation value is taken to be $v=246$ GeV. 
From the above two relations, it can be shown that the Higgs mass may
be written as: 
\begin{equation}
 m_H^2=2\mu^2=2\lambda v^2.
\end{equation}

\section{Mass Eigenstates and Mass Eigenvalues of 2 active and 2
sterile neutrino system}\label{mass_eigen}
Define the following parametrization:
\begin{equation}
 f_2=-A;f_3=-Af;f_{23}=-\epsilon Afe^{in};v'=\frac{v}{\sqrt{2}},
\end{equation}
where \begin{equation}
 \epsilon=\frac{|f_{23}|}{f_3}. 
\end{equation}

This leads to the following form of the mass-matrix: 
\begin{equation}
 M_\nu=\begin{pmatrix}
        \Delta M&M&Av'&0\\
	M&\Delta M&Af\epsilon e^{-in}v'&Afv'\\
	Av'&Af\epsilon e^{-in}v'&0&0\\
	0&Afv'&0&0
       \end{pmatrix}.
\end{equation}

Furthermore, define the parameters: \begin{equation}
                        \sigma=\Big\{1+(1+\epsilon^2)f\Big\}v'^2;\sin
a=\frac{f\epsilon v'^2 \sin n}{\rho},
                       \end{equation}
where \begin{equation*}
        \rho=\epsilon v'^2\Big[f^2(1+f^2\kappa^2+2f\kappa\cos
n)+\kappa^2
       \end{equation*}
\begin{equation}
 \hspace{10mm}+2f\kappa\cos n+2f^2k^2\Big]^{\frac{1}{2}},
\end{equation}

\begin{equation}
 \Delta M=\frac{A^2(1+f^2)v'^2\epsilon\kappa}{M}.
\end{equation}

Then the eigenstates and eigenvalues maybe written as follows: 
\begin{equation*}
 |e_1\rangle=\frac{1}{\sqrt{2}}\begin{pmatrix}
                     
e^{\frac{-ia}{2}}\\e^{\frac{ia}{2}}\\\frac{Av'}{M}(e^{\frac{ia}{2}}
+\epsilon fe^{\frac{i(2n-a)}{2}})\\\frac{Afv'}{M}e^{\frac{-ia}{2}}
                     \end{pmatrix},
\end{equation*}
\begin{equation}
  m_1=M+\frac{\sigma+2\rho}{2M}A^2.
\end{equation}
\begin{equation*}
 |e_2\rangle=\frac{1}{\sqrt{2}}\begin{pmatrix}
                     
ie^{\frac{-ia}{2}}\\-ie^{\frac{ia}{2}}\\-\frac{iAv'}{M}(e^{\frac{ia}{2
}}-\epsilon f e^{\frac{i(2n-a)}{2}})\\\frac{iAfv'}{M}e^{\frac{-ia}{2}}
                     \end{pmatrix},
\end{equation*}
\begin{equation}
 m_2=M+\frac{\sigma-2\rho}{2M}A^2.
\end{equation}
\begin{equation*}
 |e_3\rangle=\frac{1}{\sqrt{2}}\begin{pmatrix}
                               
-\frac{Afv'}{M}e^{\frac{-in}{2}}\sqrt{\frac{\sqrt{1+\epsilon^2}}{\sqrt
{1+\epsilon^2}+\epsilon}}\\\frac{Av'}{M}e^{in/2}\frac{1}{\sqrt{
1+\epsilon^2+\epsilon\sqrt{1+\epsilon^2}}}\\-e^{\frac{in}{2}}\frac{1}{
\sqrt{1+\epsilon^2+\epsilon\sqrt{1+\epsilon^2}}}\\e^{\frac{-in}{2}}
\sqrt{\frac{\sqrt{1+\epsilon^2}+\epsilon}{\sqrt{1+\epsilon^2}}}
                               \end{pmatrix},
\end{equation*}
\begin{equation}
 m_3=(\sqrt{1+\epsilon^2}+\epsilon)\frac{fv'^2}{M}A^2.
\end{equation}
\begin{equation*}
 |e_4\rangle=\frac{1}{\sqrt{2}}\begin{pmatrix}
                               
\frac{iAfv'}{M}e^{\frac{-in}{2}}\sqrt{1+\epsilon^2+\epsilon\sqrt{
1+\epsilon^2}}\\\frac{iAv'}{M}e^{\frac{in}{2}}\sqrt{\frac{\sqrt{
1+\epsilon^2}+\epsilon}{\sqrt{1+\epsilon^2}}}\\-ie^{\frac{in}{2}}\sqrt
{\frac{\sqrt{1+\epsilon^2}+\epsilon}{\sqrt{1+\epsilon^2}}}\\-ie^{\frac
{-in}{2}}\sqrt{\frac{\sqrt{1+\epsilon^2}-\epsilon}{\sqrt{1+\epsilon^2}
}}                      
\end{pmatrix},
\end{equation*}
\begin{equation}
 m_4=(\sqrt{1+\epsilon^2}-\epsilon)\frac{fv'^2}{M}A.
\end{equation}

\vspace{20mm}

\bibliography{all}

\end{document}